\documentclass[twocolumn,amsmath,amssymb]{revtex4}
\def\lsim{\raise0.3ex\hbox{$\;<$\kern-0.75em\raise-1.1ex\hbox{$\sim\;$}}}
\def\gsim{\raise0.3ex\hbox{$\;>$\kern-0.75em\raise-1.1ex\hbox{$\sim\;$}}}

\newcommand{\be}{\begin{eqnarray}}
\newcommand{\ee}{\end{eqnarray}}

\def\bea{\begin{eqnarray}}
\def\eea{\end{eqnarray}}
\usepackage{graphics}
\usepackage{epsfig}
\usepackage{array}
\usepackage{float}
\usepackage{xcolor}
\usepackage{colortbl}
\usepackage{booktabs}
\usepackage{graphicx}
\usepackage{psfrag}
\newcommand{\hp}{\hphantom0}
\usepackage{mathrsfs}

\begin{document}

\title{Quantum Cosmology with vector torsion}

\author{A. Kasem\footnote{s-ammar.kasem@zewailcity.edu.eg} and S. Khalil\footnote{skhalil@zewailcity.edu.eg}}

\affiliation{Center for Fundamental Physics, Zewail City of Science and Technology, 6 October City, Giza 12578, Egypt.}

\date{\today}

\begin{abstract}
We extend the treatment of quantum cosmology to a manifold with torsion. We adopt a model of Einstein-Cartan-Sciama-Kibble compatible with the cosmological principle. The universe wavefunction will be subject to a $\mathcal{PT}$-symmetric Hamiltonian. With a vanishing energy-momentum tensor, the universe evolution in the semiclassical and classical regimes is shown to reflect a two-stage inflationary process induced by torsion.
\end{abstract}
\maketitle

\section{Introduction}
In the theory of quantum cosmology, the universe is descried by a wavefunction subject to Wheeler–DeWitt (WDW) equation $H\Psi=0$ resulting from quantization of Einstein's General Relativity (GR) \cite{PhysRev.160.1113}. In the ensemble interpretation, the WDW wavefunction indicate the possiblity that the universe could be spontaneously created from initial quantum state and hence, the infamous classical singularity associated with the beginning of the universe can be overcome. Be that as it may, an overwhelming experimental results have been casting doubt on the standard model of cosmology based on GR \cite{Akrami:2018vks,Aghanim:2018eyx,Akrami:2018odb}. Theories with torsion have recently gained ground as a promising extension to GR that would be in line with all the observationally successful results of GR in the regime of a vanishing torsion \cite{Nikiforova:2016ngy, Valdivia:2017sat, Banerjee:2018yyi, Pereira:2019yhu, Barrow:2019bvx, Bose:2020mdm}.

Einstein-Cartan-Sciama-Kibble (ECSK) is a theory on Einstein-Cartan manifold whihch is also noteworthy for being the simplest gauge theory of gravity. It is usually thought to take over at high energy as an intermediate between GR and a more fundamental theory of quantum gravity as torsion effects are suppressed by the square of the gravitational constant ($k^2$) so it's effect is insignificant for the late universe \cite{Hehl:1976kj}. Torsion is conventionally linked to the spin because it corresponds to the spinor part in the angular momentum tensor and it is usually modeled as a macroscopically spinning fluid. However such description fails to capture the essence of the theory mainly because of its incompatibility with the cosmological principle and the shaky assumption of a classical matter at such high energies. Torsion may also emerge in different context, for instance it may arise from a gradient of scalar field \cite{Hojman:1978yz} or in sting theory \cite{Becker:2002sx}. It is also worth noting that supergravity is a natural framework where torsion, curvature and matter fields are treated in analogous manner \cite{Cremmer:1978km}. We believe a successful model should obey the cosmological principle and be studied in a quantum setting.

A study for gravity on a Riemann-Cartan manifold $U4$ in the framework of quantum cosmology has been lacking in literature. In this work, we consider the case for ECSK with vector torsion in a minimal setup. We find the corresponding WDW equation and discuss its solutions and interpretation. We find that in the absence of any other matter field, torsion can derive a period of accelerated expansion and show that it continues in the classical region.

The paper is organized as follows: In Section \ref{Setup}, we briefly describe the relevant action for quantum cosmology with vector torsion within ADM formalism. Modified WDW equation is obtained in Section \ref{WDW Section} with emphasize on the Hamiltonian being PT-symmetric which defines a physical theory with real spectrum. In Section \ref{Time interp.}, we discuss the ``clocks'' we use to interpret the solution in the semiclassicl regime. In Section \ref{Section Solu.}, we explore a model of space dominated by torsion in flat soace and show it derives inflation in the semiclassical and classical regimes. Finally, our conclusions and remarks are given in Section \ref{Conlusions}.

\section{Setup}\label{Setup}

Rieman-Cartan space time, $U_4$, in general has $40$ independent degrees of freedom; $16$ of which are encoded in the metric and $24$ in the torsion tensor\cite{Hehl:1976kj, Capozziello:2001mq}. The spatial-maximal symmetry condition reduces the metric to the FRW metric. In the ADM formalism, one splits spacetime into a space-like hypersurfaces $\Sigma$ separated by a time lapse $N(t)dt$. The FRW after the $3+1$ split and a standard choice of coordinates can be written as \cite{Arnowitt:1959ah, elbaz2012quantum}

\begin{equation}\label{FRW metric}
ds^2=-N^2dt^2+a(t)^2(\frac{1}{1-kr^2}dr^2+r^2d\Omega_2^2)
\end{equation}
on the other hand, the torsion is compatible with the same symmetries only if its traceless part vanishes and the vector and axial-vector part conform to the following form \cite{TSAMPARLIS197927};

\begin{eqnarray}
S^i_{\hp 0i}=-&S^i_{\hp i0}&=N\phi(t)\label{Vtorsion}\\
&S_{[ijk]}&=\chi(t)\label{Atorsion}
\end{eqnarray}
the choice for Axial vector Eq.(\ref{Atorsion}) minimally couples with gravity through the Lorentz-invariant measure $\sqrt{-g}d^4x$. In the following analysis we adopt the more interesting choice of a vector form Eq.(\ref{Vtorsion}) which couples as well to the metric derivative as we shall see.

Our aim is to quantize the Einstein-Hilbert action $S_{EH}=\int_{\mathcal{M}}d^4x\sqrt{-g} R$. In order to have a well-posed variational principle, one has to define boundary conditions on the manifold $\mathcal{M}$ which amounts to adding a surface term to the action. The choice of boundary does not affect the classical equations of motion. On the other hand, it directly alters its quantum counterparts owing to the differential nature of the momenta operators. Hence it is important to add the Gibbons–Hawking–York (GHY) boundary term $\int_{\partial M}2\sqrt{h}K d^3x$ to have the relevant action for quantum cosmology \cite{PhysRevLett.28.1082,PhysRevD.15.2752};
 
\begin{eqnarray}\label{Lag.4}
\mathscr{L}&=\sqrt{-g}(\hp^{(3)}R+K_{ij}K^{ij}-K^2+\mathscr{L}_{matter})
\end{eqnarray}
where the intrinsic curvature $\hp^{(3)}R$ being the Ricci scalar curvature of the surface $\Sigma$ and the extrinsic curvature tensor $K_{ij}\equiv -N\Gamma^0_{\hp ij}$. For our choice for the FRW metric and vector torsion Eqs.(\ref{FRW metric}, \ref{Vtorsion}), the intrinsic curvature and the non vanishing components of the extrinsic curvature can be calculated explicitly as follows

\begin{eqnarray}
\hp^{(3)}R&=&6\frac{k}{a^2}\\
K_{ij}&=&-g_{ij}N^{-1}(2N\phi+\frac{\dot{a}}{a})
\end{eqnarray}
putting it together, the Lagrangian Eq.(\ref{Lag.4}) reduces to

\begin{equation}\label{Lagr.}
\mathscr{L}=NKa -N^{-1}a\dot{a}^2-4\phi\dot{a}a^2-4N\phi^2 a^3-2Na^3(\rho_M+\rho_{vac})
\end{equation}

\subsection{Modified Wheeler-DeWitt equation}\label{WDW Section}

We now wish to find a modified version of WDW in light of Eq.(\ref{Lagr.}). In order to find the Hamiltonian, we first compute the conjugate momenta;

\begin{equation}
P_a=\frac{\partial\mathscr{L}}{\partial \dot{a}}=-2N^{-1}a\dot{a}-4\phi a^2\Rightarrow \dot{a}=-2N\phi a-N\frac{P_a}{2a}\label{Can. mom.}
\end{equation}
\begin{eqnarray}
P_\phi&=&\frac{\partial\mathscr{L}}{\partial \dot{\phi}}=0\\
P_N&=&\frac{\partial\mathscr{L}}{\partial \dot{N}}=0
\end{eqnarray}
hence the canonical Hamiltonian

\begin{eqnarray}
\mathscr{H}_c(P,q)&\equiv&\sum_i P_i\dot{q}_i-\mathscr{L}\nonumber \\&=&N(-\frac{P_a^2}{4a}-2\phi aP_a-Ka+2a^3(\rho_M+\rho_{vac}))\label{Classical H}
\end{eqnarray}
we note that the lapse function $N$ is a Lagrange multiplier and its equation of motion leads to the Hamiltonian constraint;

\begin{equation}
\frac{d P_N}{dt}=-\frac{\partial \mathscr{H}_c}{\partial N}\Rightarrow -\frac{P_a^2}{4a}-2\phi aP_a-Ka+2a^3(\rho+\rho_{vac})=0\label{Classical H}
\end{equation}
To quantize the system, we turn the classical observables into operators; $a\rightarrow\hat{a}$, $P_a\rightarrow -i\frac{\partial}{\partial a}$, and $\phi\rightarrow \hat{\phi}$. Then we introduce the so-called wavefunction of the universe $\Psi(a,\phi)$ which is subject to the Hamiltonian constraint $\hat{\mathscr{H}}\Psi=0$. We take into consideration the operator ordering ambiguity by taking $\hat{P}_a=-ia^q\frac{\partial}{\partial a}a^{-q}$ and $\hat{P}_a^2=-a^{-p}\frac{\partial}{\partial a}(a^p\frac{\partial}{\partial a})$. Now we obtain the wave equation for $\Psi$;

\begin{eqnarray}
a^{-p}\frac{\partial}{\partial a}(a^p\frac{\partial}{\partial a}\Psi)+&8i\phi a^2 a^q \frac{\partial}{\partial a}(a^{-q}\Psi)-4Ka^2\Psi&\nonumber\\
+&8a^4(\rho_M+\rho_{vac})\Psi=0
\end{eqnarray}
\begin{equation}
\Big( \frac{\partial^2}{\partial a^2}+(\frac{p}{a}+8i\phi a^2)\frac{\partial}{\partial a}-4K a^2+8a^4(\rho_M+\rho_{vac})\Big)\Psi=0\label{WDW BO}
\end{equation}

On physical grounds, we set $q=0$ in the last line, otherwise the equation will be plagued by an imaginary effective potential and hence the Hamiltonian would be non-Hermition. Having done that, we are still lift with a non-Hermition operator however it is now $\mathcal{PT}$-symmetric. For a Hamiltonian endowed with such symmetry, on can safely define a physical theory with real spectrum, unitary time evolution, and states of positive norm under a $\mathcal{CPT}$ inner product \cite{Bender_2007};

\begin{equation}
\langle f\mid g\rangle=\int_C dx [\mathcal{CPT}f(x)]g(x)
\end{equation}
We however are not interested in the time evolution or the rest of the energy spectrum as WDW dictates $\hat{H}\Psi=0$ so we are only interested in the zero energy state.

\subsection{Probability current and emergent time}\label{Time interp.}

We follow the semiclassical approach in tackling the infamous problem of time. We rewrite the wave function in the WKB ansatz as $\Psi(a,\phi)=R(a,\phi)e^{iS(a,\phi)}$; where $R(a,\phi)$ and $S(a,\phi)$ being real functions. Eq.(\ref{WDW BO}) can be cast into

\begin{equation}
\frac{p}{4a^2} \frac{R^\prime}{R}\hbar^2+\frac{1}{4a}\frac{R^{\prime\prime}}{R}\hbar^2-\frac{{S^\prime}^2}{4a}-2a\phi S^\prime-Ka+2a^3(\rho_M+\rho_{vac})=0\label{HJ}
\end{equation}
\begin{equation}
(8a^3\phi+2aS^\prime)\frac{R^\prime}{R}+p S^\prime+aS^{\prime\prime}=0
\end{equation}
Although we are working with natural units, we temporarily restored the $\hbar$ dependence to show that up the zeroth order of $\hbar$, Eq.(\ref{HJ}) is the Hamilton-Jacobi equation for the action $S$. Upon comparison with its classical counterpart Eq.(\ref{Classical H}), we can relate the function $S$ to the canonical momentum $P_a=\frac{\partial S}{\partial a}$. Using the canonical momentum from Eq.(\ref{Can. mom.});

\begin{equation}\label{QHJ}
P_a=\frac{\partial\mathscr{L}}{\partial \dot{a}}=-2a\dot{a}-4\phi a^2=S'
\end{equation}
this guidance equation will be useful to establish a bridge between the quantum theory and a classical theory where the scale factor evolves in time. Now we attempt to find the probability current for the WKB wavefunction in a standard calculation

\begin{eqnarray}
\Psi^* H\Psi&=&\Psi H^\dagger \Psi^*\nonumber\\
\Psi^*\frac{\partial^2\Psi}{\partial a^2}+\Psi^*(8i a^2\phi+\frac{p}{a})\frac{\partial\Psi}{\partial a}
&=&\frac{\partial^2\Psi^*}{\partial a^2}\psi-\frac{\partial\Psi^*}{\partial a}(8i a^2\phi-\frac{p}{a})\psi\nonumber
\end{eqnarray}
\begin{eqnarray}
\frac{\partial}{\partial a}\bigg(a^{p}\Big(\Psi^*\frac{\partial\Psi}{\partial a}-\frac{\partial\Psi^*}{\partial a}\psi\big)+8i\phi a^{p+2}\psi^*\psi\bigg)&=&0\label{current cons}
\end{eqnarray}
this equation implies the conservation of a current $\partial_aJ_a=0$ which can be written under the WKB ansatz as;

\begin{equation}
J_a\equiv a^p(\Psi^*\frac{\partial\Psi}{\partial a}-\Psi\frac{\partial\Psi^*}{\partial a})+8i\phi a^{p+2}\psi^*\psi=2iR^2a^{p}(S'+4a^2\phi)
\end{equation}
from the conservation Eq.(\ref{current cons}) and by substitution from the guidance equation Eq.(\ref{QHJ}), one gets an equation relating the probability density to the time evolution of the scale factor

\begin{equation}\label{prob den}
R^2=\frac{a^{-p}}{2a\dot{a}}C_0(\phi)^2=\frac{C_0(\phi)^2}{2a^{p+2}H(a)}
\end{equation}
where $C_0(\phi)$ is an arbitrary function and we write it squared for convenience. In this regime, one defines an emergent semiclassical time by solving the WDW for the time independent probability density $R(a,\phi)$ and solve Eq.(\ref{prob den}) for the evolution of a classical $a(t)$. It is interesting that for a finite probability density and Hubble parameter in the early universe $a\rightarrow 0$, the ordering parameter has to be $p=-2$ \cite{He:2015wla}. We adopt this choice in the following analysis.

\section{Solution in Flat Space}\label{Section Solu.}

\begin{figure}[t]
\centering
\includegraphics[width=0.5\textwidth]{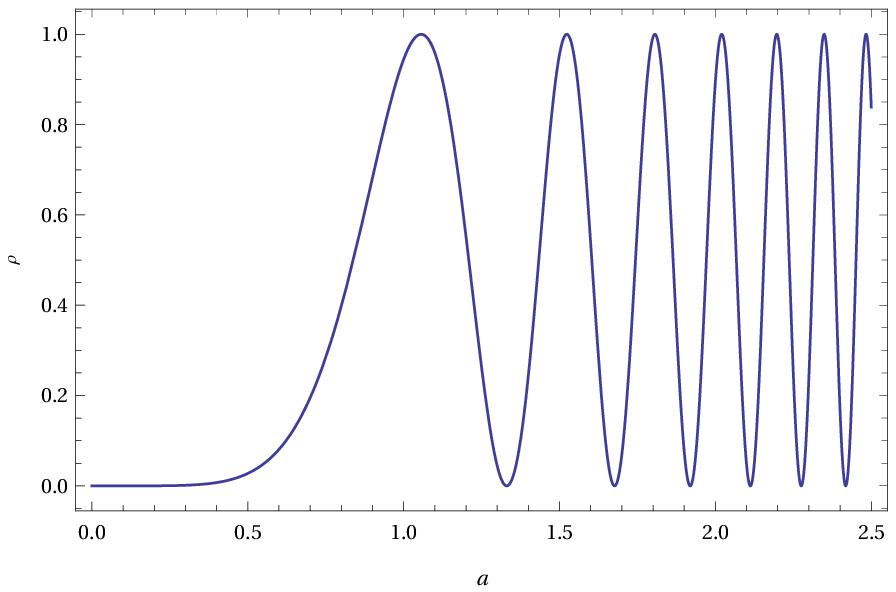}
\includegraphics[width=0.5\textwidth]{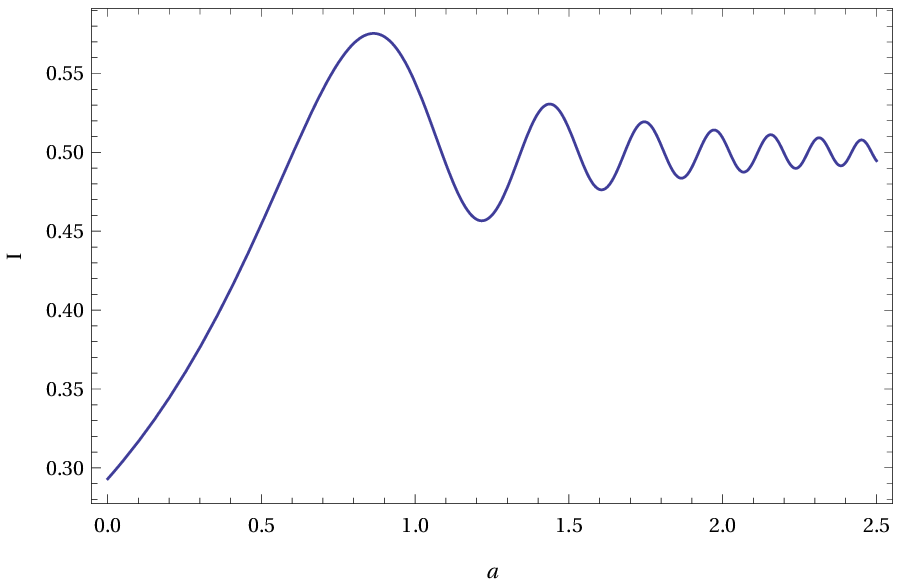}
\caption{Probability density as a function of scale factor and the integral $I$ at $\phi=1$}\label{rhoVa}
\end{figure}

Now, we attempt to examine a minisuperspace model with an ordering parameter $p=-2$ and in flat space $K=0$. Even with all forms of energy vanishing, the torsion gurantees a nontrivial solution to the wavefunction. Namely, Eq.(\ref{WDW BO}) and its solution take the form

\begin{equation}
\Big( \frac{\partial^2}{\partial a^2}+(8i\phi a^2-\frac{2}{a})\frac{\partial}{\partial a}\Big)\Psi(a,\phi)=0
\end{equation}

\begin{eqnarray}
\Psi(a,\phi)&=\nonumber C_1(\phi)\frac{e^{-\frac{8}{3}ia^3\phi}-1}{\phi}+C_2(\phi)
\end{eqnarray}
We demand that the universe has a zero probability of emerging in a singular state; this is equivalent to setting a boundary condition $\Psi(0,\phi)=0\Rightarrow C_2(\phi)=0$. We note that the variable $\phi$, by construction, is not dynamical hence we cannot determine the particular form of the function $C_1(\phi)$. However at a constant $\phi$, it can be treated as a normalization constant. Now we write the wavefunction as;

\begin{equation}
\Psi(a,\phi)= \frac{C_1(\phi)}{\phi}(e^{-\frac{8}{3}ia^3\phi}-1)
\end{equation}
The quantum regime is defined for a small scale factor $a<1$, and a large field $\phi\approx 1$. In this regime, the wavefunction represents an ensemble of universes that would require an external observer to make sense of any excpectation value. However when the universe is large enough and the field is very small ($a>1$, $\phi\ll 1$), it can interact with itself and evolves in a classical trajectory. The wavefunction in that limit would be $\Psi\sim a^3$ and one can immediately get the evolution using Eq.(\ref{QHJ}) with $S^\prime=0$ to yield;

\begin{equation}
\frac{\dot{a}}{a}=-2\phi\Rightarrow a=a_0e^{-2\int_{t_0}^t\phi(t)dt}
\end{equation}
this result agrees with the result from the classical analysis \cite{Kranas:2018jdc}. In an earlier stage the universe can enter the classical domain $a>1$ with a large value for the torsion field $\phi\approx 1$, and the probability density $\rho\equiv\mid\Psi\mid^2=\frac{C_1(\phi)^2}{\phi^2}\sin(\frac{4}{3}a^3\phi)^2$ is used in the dynamical interpretation c.f. Eq.(\ref{prob den}) to signify the universe evolution from the state $a_1$ to $a_2$ as $\int_{a1}^{a_2}\rho da\sim \int_{t_1}^{t_2} a dt$.

We note from fig.(\ref{rhoVa}) that there is a zero probability of finding the universe at certain states $a_n=(\frac{3}{4\phi}n\pi)^{1/3}$ for integer $n$. And despite the nonrenormalizability, the quantity $I\equiv\frac{1}{\Delta a}\int_a^{a+\Delta a}\rho da$ as depicted from fig.(\ref{rhoVa}) remains regular and would eventually converge to a particular value as $a$ increases and $\Delta a$ tends to zero;

$$\rho(a\gg 1)\equiv\lim_{\substack{\Delta a\rightarrow 0 \\ a\gg 1}}\frac{\int_{a}^{a+\Delta a}\rho da}{\Delta a}=\frac{C_1(\phi)^2}{2\phi^2}$$
so in this regime the probability density only depends on $\phi$ and using Eq.(\ref{prob den}), we can get the classical evolution of the scale factor $a$

\begin{equation}
H(a\gg 1,\phi\sim 1)=\frac{C_0^2}{C_1^2}\phi^2\Rightarrow a(t)=a_0e^{\int_{t_0}^t \frac{C_0^2}{C_1^2}\phi^2 dt}
\end{equation}
because $\phi$ should be treated as a quantum field in this region, the answer depends on the functions $C_1(\phi)$ and $C_0(\phi)$ which can be founded in theory by quantizing the matter field responsible for the torsion effect.

\section{Conclusions}\label{Conlusions}
We have investigated a universe with a nonvanishing torsion in its early stages. The corresponding WDW equation was found and shown to be $\mathcal{PT}$-symmetric. The ordering factor of $-2$ was adopted and we showed the universe undergoes a two-stage inflationary process under the influence of torsion in the absence of conventional matter fields that source the energy-momentum tensor.

We have so far been treating the torsion as a built-in function of time embedded in the geometry. However such function must be sourced by a matter field. It is an interesting avenue to explore the evolution of different matter fields allowing such form of torsion and study its effect on structure formation. Recently possible Dirac and Maxwell fields allowing forms of torsion compatible with the cosmological constant has been discussed \cite{Cabral:2020mzw}. Also we have in a previous work considered the triad field in classical regime and it was shown to derive an exponential inflation \cite{Kasem:2020ddi}.

\bibliographystyle{unsrt}
\bibliography{refs}

\end{document}